\documentclass[sort,compress]{elsarticle}
\usepackage[utf8]{inputenc}
\usepackage[english]{babel}
\usepackage{amssymb,amsmath,amsthm}
\usepackage{amsfonts}
\usepackage{euscript}
\usepackage{hyperref}
\allowdisplaybreaks[4]

\newcommand{\RO}[1]{\EuScript{#1}}

\newtheorem{thm}{Theorem}
\newtheorem{proposition}{Proposition}
\theoremstyle{remark}

\begin{document}
\title{Infinitely Many Commuting Nonlocal Symmetries for Modified Mart\'\i{}nez Alonso--Shabat Equation}

\author{Hynek Baran}
\ead{Hynek.Baran@math.slu.cz}
\address{Mathematical Institute, Silesian University in Opava, \\Na  Rybn\'{\i}\v{c}ku 1, 746 01 Opava, Czech Republic}

\begin{abstract}
We study the modified Mart\'\i{}nez Alonso--Shabat equation
$$u_y u_{xz} + \alpha u_x u_{ty}  - (u_z + \alpha u_t)u_{xy} = 0$$ and present its
recursion operator and an infinite commuting hierarchy of full-fledged nonlocal
symmetries. To date such hierarchies were found only for very few integrable systems in more than three independent variables.
\end{abstract}


\maketitle

\section{Introduction}

Integrable systems are well known to have an important role in modern mathematical physics,
see e.g.\
\cite{ASF, Olver:1993, KVV, Serg:2017fs}.
An important feature of integrable partial differential systems is that any such system belongs to an infinite hierarchy of pairwise compatible systems that can be seen as symmetries of each other, cf.\ for example \cite{ASF, Serg:2017ir, KVV, df}. Such an infinite hierarchy of symmetries is an important sign of integrability and, on the other hand, a useful structure attached to a given integrable system, as such a hierarchy provides, to an extent, the structure behind infinite families of explicit exact solutions like multisolitons, cf.\ e.g.\ the discussion in \cite{ASF,  Olver:1993}; see also \cite{Olver:1993, KVV, df, Bruzon} for applications of symmetries in general.

For integrable partial differential systems in more than two independent variables the symmetries in question, as well as the conservation laws, are typically nonlocal,
see e.g.\ \cite{ASF, KVV, Serg:2017ir, Serg:2017fs,Baran:2016hd, Baran:2018cv},
which makes the task of finding their commutation relations quite difficult, cf.\ e.g.\ \cite{KVV,Baran:2018cv}. There is a technique \cite{Serg:2008, Morozov:2014ff} allowing one to find an infinite hierarchy of nonlocal symmetries and establish its commutativity using a Lax pair of the system under study for a fairly broad class of integrable multidimensional systems with isospectral Lax pairs involving an essential parameter. Given the importance of such hierarchies, as discussed above, it is natural to check whether indeed more examples of hierarchies of commuting nonlocal symmetries can be found using this technique. In the present paper we show that this can be done for the modified Mart\'\i{}nez Alonso--Shabat equation in four independent variables (4D) and present an infinite commutative hierarchy of full-fledged nonlocal symmetries for this equation as well as a recursion operator.\looseness=-1

\section{Modified Mart\'\i{}nez Alonso--Shabat equation}
Consider the modified Mart\'inez Alonso--Shabat equation \cite{Morozov:2014ff}
\begin{equation}\label{TheEq0}
u_y u_{xz} + \alpha u_x u_{ty}  - (u_z + \alpha u_t)u_{xy} = 0
\end{equation}
involving  a nonzero real parameter $\alpha$.

Equation (\ref{TheEq0}) is an integrable 4D PDE as it
has \cite{Morozov:2014ff} a Lax pair involving the spectral parameter $\lambda\neq\alpha$
\begin{equation} \label{LaxPair}
r_y = \frac{\lambda}{\alpha} \frac{u_y}{u_x} r_x, \quad r_z =\frac{\lambda}{\alpha}  \frac{u_z + \alpha u_t}{u_x} r_x - \lambda r_t,
\end{equation}
cf.\ e.g.\ \cite{KVV, Serg:2017ir, Serg:2017fs} and references therein
for integrable 4D systems in general.

Identifying $z$ and $t$ in (\ref{TheEq0}) yields \cite{Morozov:2014ff} a 3D integrable reduction of the latter,
\begin{equation}\label{abc}u_y u_{tx}-(\alpha+1)u_t u_{xy}+\alpha u_x u_{ty}=0.\end{equation}

In turn, (\ref{abc}) is, up to a possible relabeling of independent variables and multiplication by an overall constant, nothing but the Veronese web equation, also known as the the ABC equation,
\begin{equation}\label{abceq}A u_x u_{ty} +B u_y u_{tx} + C u_t u_{xy} =0,\quad A+B+C=0,\end{equation}
which describes three-dimensional Veronese webs and is a subject of intense research,
see e.g.\ \cite{Kumar,KMV} and references therein.
Thus, (\ref{TheEq0}) can be seen as a natural 4D generalization of (\ref{abc}) and hence of (\ref{abceq}).

To simplify further computations, in what follows we shall work with equation (\ref{TheEq0}) in the form 
\begin{equation} \label{TheEq}
u_{ty} = \frac{\alpha  u_t u_{xy} + u_z u_{xy} - u_y u_{xz}}{\alpha  u_x}
\end{equation}
solved for $u_{ty}$.


\section{The recursion operator} 

Starting with (\ref{LaxPair}) and using the deformation procedure described in \cite{Serg:2017wl} (cf.\ also \cite{Serg:2017ir}) we readily find that (\ref{TheEq}) admits, in addition to (\ref{LaxPair}),
a Lax pair
\begin{equation}\label{cov_q}
\begin{split}
q_y &=\frac{ \lambda u_y q_x  + (\alpha-\lambda) q u_{xy} }{\alpha  u_x},\\
q_z &=\frac{\lambda\left((\alpha u_t + u_z) q_x   - \alpha u_x q_t  - q u_{xz}\right) + \alpha  q u_{xz}}{\alpha  u_x}.
\end{split}
\end{equation}

In particular, for any given $\lambda$ equations (\ref{cov_q}) define a covering, which we denote by $\mathcal{Q}_\lambda$, over (\ref{TheEq}); see e.g.\ \cite{KVV} for general background on coverings.

Unlike $r$, if $q$ satisfies (\ref{cov_q}) then it is a nonlocal symmetry shadow for (\ref{TheEq}) in the covering $\mathcal{Q}_\lambda$, i.e., roughly speaking, $\varphi=q$ satisfies the linearized version  of (\ref{TheEq}),\looseness=-1
\begin{multline}  \label{lin}
\ell_\mathcal{E}(\varphi) \equiv
\frac{
  u_t u_{xy} D_{x} (\varphi)
- u_x u_{xy}  D_{t} (\varphi)
+ u_x^2 D_{ty}    (\varphi)
- u_t u_x D_{xy}  (\varphi)
}
{u_x^2}    \\*
+
\frac{
u_z u_{xy} D_{x}        (\varphi)
- u_y u_{xz} D_{x}        (\varphi)
+ u_x u_{xz} D_{y}        (\varphi)
- u_x u_{xy} D_{z}        (\varphi)
}
{\alpha  u_x^2}    \\*
+
\frac{
  u_x u_y D_{xz}          (\varphi)
- u_x u_z D_{xy}          (\varphi)
}
{\alpha  u_x^2}
= 0
\end{multline}
modulo (\ref{TheEq}), (\ref{cov_q}) and differential consequences thereof.

Here $D_x$, $D_y$ etc.\ denote total derivatives in the appropriate covering  over (\ref{TheEq}), e.g.\ $\mathcal{Q}_\lambda$ for $q$, cf.\ e.g.\ \cite{KVV} for relevant definitions.

Following \cite{Serg:2017wl, Serg:2017ir},
upon formally replacing $\lambda q$ by $\varphi$ and $q$ by $\psi$ in (\ref{cov_q}), we readily arrive at the following
\begin{proposition}\label{p1}
Equation (\ref{TheEq})  admits a recursion operator $\RO{R}$ defined by the  relations
\begin{equation}\label{TheRO}
\begin{split}
\psi_y &= \frac{u_y \varphi_x -  u_{xy} \varphi + \alpha   u_{xy}  \psi}{\alpha  u_x},\\
\psi_z &= \frac{(\alpha  u_t + u_z)\varphi_x - \alpha  u_x \varphi_t - u_{xz} \varphi +  \alpha  u_{xz} \psi}{\alpha  u_x},
\end{split}
\end{equation}
meaning that for any nonlocal symmetry shadow $\varphi$ for (\ref{TheEq})  $\RO{R}$  produces another nonlocal symmetry shadow $\RO{R}(\varphi)\stackrel{\mathrm{def}}{=}\psi$ for (\ref{TheEq}).
\end{proposition}

In other words, the above $\RO{R}$ defines a B\"acklund auto-transformation for the linearized version (\ref{lin}) of (\ref{TheEq}),
see e.g.\ \cite{Mar, KVV, Serg:2017ir, pap20, Serg:2005} and references therein
for details on this approach to recursion operators.

While using $\RO{R}$ one readily can construct infinite hierarchies of nonlocal symmetry shadows for (\ref{TheEq}), this leaves one with the problem of finding a (minimal) covering in which all these shadows could be lifted to full-fledged nonlocal symmetries of (\ref{TheEq}), since only for those one can rigorously establish their commutation relations.

In what follows we shall take a slightly different route, using (\ref{cov_q}) rather than $\RO{R}$, to produce just such a hierarchy of nonlocal symmetries for (\ref{TheEq}) and establish their commutativity.


\section{Nonlocal symmetries}

While, as we have seen in the preceding section, $q$
is a nonlocal symmetry shadow in the covering $\mathcal{Q}_\lambda$, this shadow cannot be lifted to a full-fledged nonlocal symmetry in the covering under study.

To circumvent this difficulty, consider a formal expansion $q=\sum_{i=0}^\infty q_i\lambda^i$.
Substituting this expansion into (\ref{cov_q}) shows that
$q_0=F u_x$,
where $F(x,t)$ is an arbitrary function,
while the remaining $q_i$ are defined by the equations
\begin{equation*}
\begin{split}
{(q_1)_y} &= \frac{ \alpha u_{xy} {q_1} + ( u_{xx} u_y -  u_{xy} u_x) F +   u_x u_y F_x}{\alpha  u_x},\\
{(q_1)_z}  &=
%
\frac{
\alpha  u_{xz} q_1+
 \left(
  \alpha (u_{t} u_x)_x
  +  u_{xx} u_z
  -  u_{xz} u_x
  \right) F
+   ( \alpha u_t + u_z)  u_x F_x
- \alpha  u_x^2 F_t
}{\alpha  u_x}, \\
%
%
{(q_i)_y} &= \frac{\alpha u_{xy} {q_i} - u_{xy} {(q_{i-1})} + u_y {(q_{i-1})}_x}{\alpha  u_x},\\
{(q_i)_z} &= \frac
	{\alpha u_{xz} {q_i} - u_{xz} {(q_{i-1})} - \alpha  u_x {(q_{i-1})}_t + \alpha  u_t {(q_{i-1})}_x  +  u_z {(q_{i-1})}_x}
	{\alpha  u_x}, 
\end{split}
\end{equation*}
$ i=2,3,\dots$, that define an infinite-dimensional covering, which we denote by
$\mathcal{Q}_\infty,$ over (\ref{TheEq}). 

\begin{thm}\label{t1}
Infinite prolongations of the vector fields
\begin{equation}\label{nls}
Q_i=\displaystyle q_i\frac{\partial}{\partial u}+\sum\limits_{j=1}^{\infty} B_{i}^{j}\frac{\partial}{\partial q_j},\quad i=1,2,\dots,
\end{equation}
form an infinite hierarchy of commuting nonlocal symmetries for (\ref{TheEq}) in the covering
$\mathcal{Q}_\infty$.

Here
\begin{multline}\label{b}
B_{i}^{j}=
%
 \frac{ \left( \left[ u_x, q_{{i+j-1}} \right]_x
				 - \alpha \left[ u_x, q_{{i+j}} \right]_x \right) F
        -  \left( \alpha\,q_{{i+j}}-q_{{i+j-1}} \right) u_x \, F_x
 } {\alpha\,u_x}
\\*
  + \frac{ \left( q_{{i+j-s(i,j)-1}} \right)_x q_{{s(i,j)+1}}
 }{u_x}+ \sum _{k=1}^{s(i,j)}
 				\frac{\alpha \left[q_{i+j-k},  q_k \right]_x
 					  - \left[q_{i+j-k-1},  q_k \right]_x}
				{ \alpha u_x},
\end{multline}
$s(i,j)=\min \left( i-1,j-1 \right)$ and $[A,B]_x = A_x B - A B_x$.
\end{thm}

Before proceeding to the proof of the theorem note that 
by the very construction we have $q_{i+1}=\RO{R}(q_i)$, so the commutativity of infinite prolongations of $Q_i$ suggests that 
the above recursion operator $\RO{R}$ could be hereditary (cf.\ e.g.\ \cite{KVV,Olver:1993} and references therein on the hereditary property in general), at least when restricted to the span of shadows $q_i$, $i=1,2,\dots$, which could provide some additional insight into how the hereditary property works in the multidimensions.

\noindent{\em Proof.}
First of all, it is immediate that $q_i$ is a nonlocal symmetry shadow for (\ref{TheEq}) for all $i=1,2,\dots$ since so is $q$.

Inspired by \cite{Serg:2008,Morozov:2014ff}, we were able to find the lifts of $q_i$, $i=1,2,\dots$,
to the covering $\mathcal{Q}_\infty$. These lifts are  nonlocal symmetries $Q_i$ for (\ref{TheEq}) given by (\ref{nls}).

Now, commutativity of the infinite prolongations of $Q_i$ is easily seen (cf.\ \cite{Serg:2008,Morozov:2014ff}) to be tantamount to that of the flows
\begin{equation}\label{fqi}
\partial u/\partial\tau_i=q_i,\quad \partial q_i/\partial\tau_j=B_{i}^{j},\quad i,j=1,2,\dots
\end{equation}
i.e., to the requirement that the relations
\begin{equation}\label{cfij}
\hspace*{-5mm}
\partial^2 u/\partial{\tau_i}\partial\tau_j=\partial^2 u/\partial{\tau_j}\partial\tau_i,\ \
\partial^2 q_k/\partial{\tau_i}\partial\tau_j=\partial^2 q_k/\partial{\tau_j}\partial\tau_i,
\ \
i,j,k=1,2,\dots,\hspace{-5mm}
\end{equation}
hold by virtue of (\ref{TheEq}) and (\ref{fqi}) and their differential consequences,
which in turn is readily verified by straightforward but tedious computation.
$\Box$

In closing note that finding explicit form of the generators and providing rigorous proofs of commutation relations for infinite-dimensional algebras of nonlocal symmetries  for multidimensional integrable PDEs, rather than merely finding shadows of nonlocal symmetries, appears to be quite rare, especially in the case of four (or more) independent variables. In particular, there are only a few earlier examples known to the present author where this was achieved in 4D, namely, the commutative hierarchies of nonlocal symmetries for the self-dual Yang--Mills equations \cite{ACT} and for the Mart\'\i{}nez Alonso--Shabat equation \cite{Morozov:2014ff}. Interestingly, the situation appears to be quite different in 3D, where infinite-dimensional noncommutative algebras of nonlocal symmetries for a number of dispersionless integrable systems were found by direct computations, see e.g.\ \cite{KMV,FM,Baran:2016hd,Baran:2018cv}.\looseness=-1



\section*{Acknowledgments}
This research was supported by the grant IGS/11/2019 of Silesian University in Opava.
I would like to thank Artur Sergyeyev for helpful advice.
Many symbolic computations in the paper were performed using the software \texttt{Jets}~\cite{jets}.

{\setlength{\bibsep}{0.3pt}
\bibliographystyle{elsarticle-num} 
{\footnotesize \bibliography{\jobname}}}

\end{document}